\documentclass{elsart}
\usepackage{epsfig}
\usepackage{amssymb}
\journal{INPC2004}
\def\be{\begin{eqnarray}}
\def\ee{\end{eqnarray}}
\def\lsim{\mathrel{\rlap{\lower3pt\hbox{\hskip1pt$\sim$}}
     \raise1pt\hbox{$<$}}} %less than or approx. symbol
\def\gsim{\mathrel{\rlap{\lower3pt\hbox{\hskip1pt$\sim$}}
     \raise1pt\hbox{$>$}}} %greater than or approx. symbol

\begin{document}

\runauthor{Brown, Lee \& Rho}

\begin{frontmatter}
\title{What Hath RHIC Wrought ? : The Chiral Restoration Phase
Transition Found at RHIC}

\author[suny]{Gerald E. Brown,}
\author[pnu]{Chang-Hwan Lee}
\author[saclay]{and Mannque Rho}

\address[suny]{Department of Physics and Astronomy,\\
               State University of New York, Stony Brook, NY 11794, USA \\
	       (\small E-mail: Ellen.Popenoe@sunysb.edu)}

\address[pnu]{Department of Physics, 
%     and Nuclear physics \& Radiation technology Institute (NuRI),\\
      Pusan National University,
          Pusan 609-735, Korea\\ (E-mail: clee@pusan.ac.kr) }
\address[saclay]{Service de Physique Th\'eorique, 
  CEA Saclay, 91191 Gif-sur-Yvette
     c\'edex, France\\
\& Department of Physics, Hanyang University, Seoul 133-791, Korea\\
     (E-mail: rho@spht.saclay.cea.fr)}
%\thanks[geb]{Ellen.Popenoe@sunysb.edu}
%\thanks[chl]{clee@pusan.ac.kr}
%\thanks[rho]{rho@spht.saclay.cea.fr}

\begin{abstract}
Given Brown/Rho scaling\cite{br91} 
we thought of the mesons which are observed in
Nambu-Jona Lasinio theory as collective modes. The $\pi, \sigma, \rho$
and $A_1$, should go smoothly through $T_c$ as the temperature rose,
going massless as $T$ comes to $T_c$ \cite{BBP91,BJBP93}.
\end{abstract}
%\begin{keyword}
%\PACS{97.60.Lf; 97.80.Jp}
%\end{keyword}

\end{frontmatter}

%\renewcommand{\thefootnote}{\fnsymbol{footnote}}
%\setcounter{footnote}{0}
%%%%%%%%%%%%%%%%%%%%%%%%%%%%%%%%%%%%%%%%%%%%%%%%%%%%%%%%%%%%%%%%%%%%%%%%%%%%%%%%

%\section{Introduction\label{intro}}

In a general sense, this is what happens; namely it can be seen
directly from the increase in entropy above $T_c$ in lattice
gauge simulation (LGS) that there are 32 essentially massless excitations
just above $T_c$, the number of degrees of freedom in the mesons
denumerated above, together with their isospin partners
(There is no dependence on isospin above $T_c$.)
These 32 degrees of freedom are called instanton molecules;
equivalently, chirally restored mesons.

The most straightforward way to construct the chiral restoring
phase transition is to make the $\pi$ and $\sigma$ masses,
in the chiral limit, zero on both sides of $T_c$; i.e., to carry
the $\pi$ and $\sigma$ mesons smoothly through $T_c$ since the
transition, from the point of view of the mesons which dominate RHIC
physics, is second order. Here the first
surprise was encountered. LGS showed the quark and antiquark masses
to be $\gsim 1 GeV$ above $T_c$. That means that to make the $\pi$
and $\sigma$ massless, they must be bound by an attractive interaction
which is strong enough to provide 2 GeV binding energy.

Lattice calculations showed that the Coulomb (color singlet) interaction
increased substantially as $T$ went above $T_c$.
(In response to the chiral symmetry breaking order parameter
of $4\pi f_\pi\sim 1$ GeV, this parameter, going to zero and the 
thermodynamic variables, the meson masses being zero, the gauge
coupling jumps back towards the infrared at $T_c$. Just above $T_c$
the attractive Coulomb interaction plus the Ampere's law velocity-velocity
interaction which together give a color singlet gauge coupling
of $\alpha_s=1$,
bind the quark and antiquark by $\sim 0.5$ GeV, 1/4
of the way down to zero from the 2 GeV sum of quark and antiquark
thermal masses.

In fitting the temperature dependence of the melting of the soft glue,
Brown et al. \cite{BGLR} could fit the curve of the LGS
by a four-point Nambu-Jona Lasinio interaction, which has only
two parameters, the coupling $G$ which includes all information
about gluonic interactions, and the cut off $\Lambda$.
The surprising result was that $G$ decreased by only $6\%$ in going
through chiral restoration at $T_c$. Thus, one had the strength
of the NJL above $T_c$, little changed from below $T_c$.
($G$ comes mainly from the 't Hooft interaction which decreases
only little at $T_c$.)

Now the Coulomb bound states, the so called Furry representation,
furnished a representation of completely degenerate quark-antiquark
bound states. The Nambu-Jona Lasinio connected all of these
states, with equal, attractive matrix elements. The problem is
just that of Brown's giant dipole collective state except
that the $\pi$ and $\sigma$ collective modes are those of Goldstone
modes, or Anderson modes in a superconductor. They must be of zero
mass at $T_c$ if a smooth, essentially second order except for the
few baryons, transition is to be made at RHIC, so we are guided
by the chiral symmetry.

The way in which these modes are made massless is simple, but
instructive. Namely, since the quark and antiquark each have large
thermal masses of $\sim 1$ GeV, they must be brought down to zero
mass by very strong attractive interactions, with coupling constants
$g_{\rm eff} \sim 7$. Because of this the massless chirally restored
mesons are extremely tightly bound, and summing quark-antiquark
states to make them into vibrations brings them down in radius to
$\sim 0.1$ fm.

With increasing temperature above $T_c$, roughly at $1.5 T_c$, Debye
screening sets in and the Coulomb attraction weakens. Also, as the
``epoxy" (hard glue) melts, the NJL four-point interaction weakens
and at $T\lsim 2 T_c$ the bound states break up.
For temperatures $\gsim 2 T_c$ the perturbative nature probably
sets in, but RHIC does not go more than $\sim 2 T_c$.

The early universe went through the phase of $\bar q q$-bound states,
essentially chirally restored mesons, $\sim 10$ microseconds following
the big bang.

\begin{figure}
\centerline{\epsfig{file=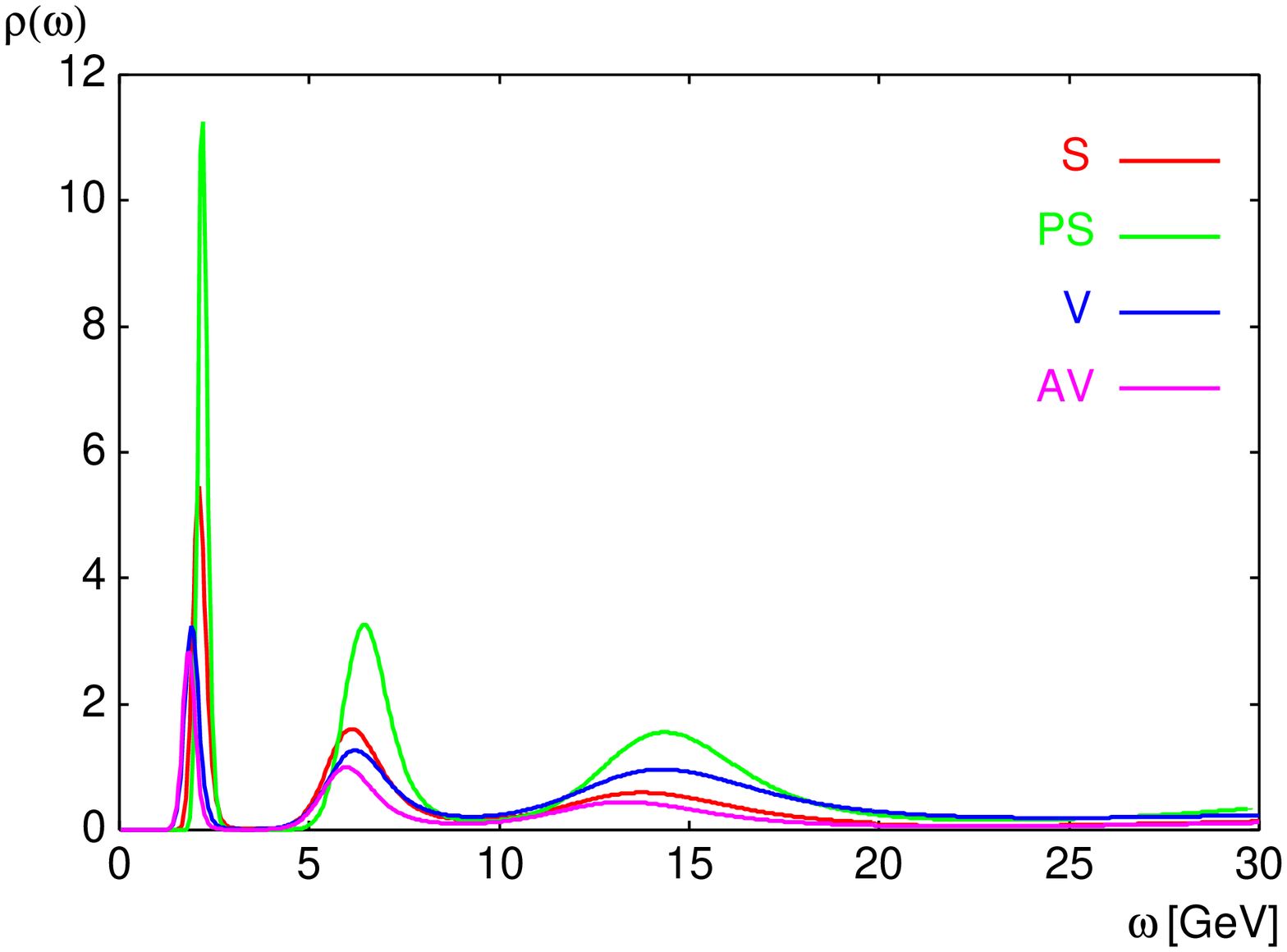,height=2in}
\epsfig{file=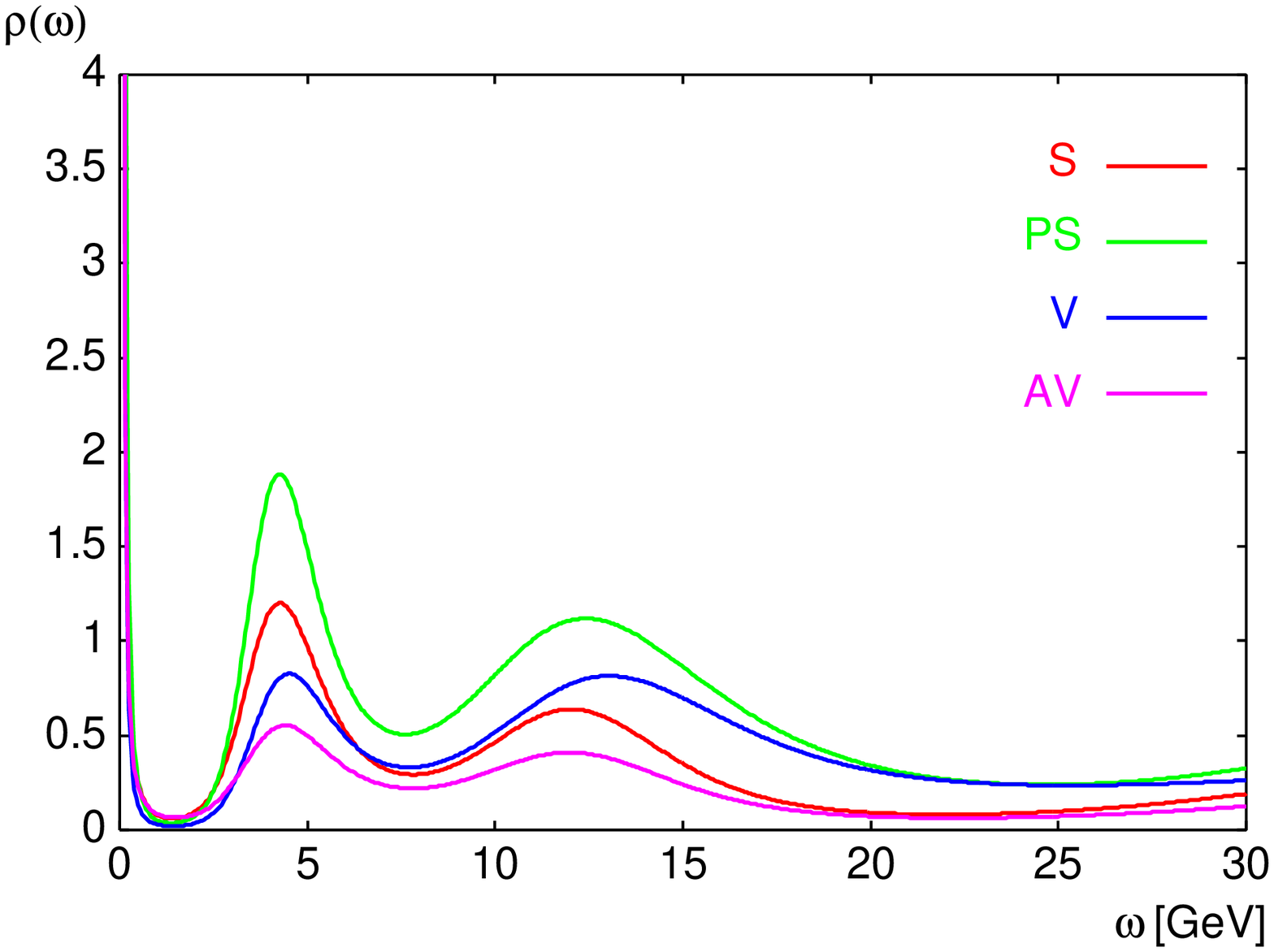,height=2in}}
\caption{Spectral functions of Asakawa et al.\cite{asakawa03}.
Left panel: for $N_{\tau}=54$ ($T\simeq 1.4 T_c$).
Right panel: for $N_{\tau}=40$ ($T\simeq 1.9 T_c$).
In each graph only the lowest vibrations is physical; 
the higher ones are lattice artifacts.
Also these lattice results are only for heavy quarks.
With the additional (attractive) interactions noted above
the lowest vibration at $T=1.4 T_c$ moves down nearly
to zero energy.}
\label{fig1}
\end{figure}

One of the best features of the chirally restored mesons is that they
are seen in lattice gauge simulations, as giant resonances\cite{asakawa03},
very similar to the giant dipole resonance in nuclear physics. 
We show these results in Fig.~\ref{fig1}.
They have a complete SU(4) symmetry, as described by the theory, scalar,
pseudoscalar, vector and axial vector excitations being degenerate.
There is no isospin dependence. The chirally restored mesons are collective
vibrations.

The change in constituent behavior is particularly interesting in going
from below $T_c$ to above $T_c$. As $T$ goes up to $T_c$ from below,
the rho meson mass goes to zero in the chiral limit\cite{HY:PR},
also the $g_V^\star\rightarrow 0$. The $\rho$ decouples from the
pions. The equation of state of the large number of massless,
weakly interacting particles becomes very soft.

In going above $T_c$ the interactions become very strong, strongest just
above $T_c$ where the masses of the chirally restored mesons are brought
to zero (in the chiral limit). Above $T_c$, with chiral restoration,
the $\rho$ and $A_1$ are equivalent as are the $\pi$ and $\sigma$.

As noted earlier the interactions show no isospin dependence. Since
there are only two different species, $\pi$ and $\rho$,
nonlinearity in the vibrations has to come from the reaction
\be
\rho\leftrightarrows 2\pi.
\ee
This nonlinearity has been estimated \cite{BLR},
resulting in a width $\Gamma (\rho\rightarrow 2\pi)$
estimated as $\sim 380$ MeV. The region of temperature just above
$T_c$ is, therefore, the most strongly interacting region, and
chemical equilibration of the various species is assured.
As the temperature decreases from above $T_c$ to below $T_c$
in the expansion of the fireball in RHIC, the hot material goes
from the strongly interacting region in which it is equilibrated into
the very weakly interacting region below $T_c$. Thus, chemical freezeout
comes as the system goes from $T_c+\epsilon$ down to $T_c-\epsilon$,
the band of energies which are mixed by the explicit chiral symmetry
breaking, $\epsilon\sim 5 $ MeV.

The chemical freezeout occurs with the hadrons off-shell. (Just above
$T_c$ they are dynamically bound into colorless chirally restored
mesons.) Whereas most of the abundances which result from chemical
freezeout at $T_c-\epsilon$, with the hadrons considered
as on-shell, are the same as if the hadrons are equilibrated
off-shell above $T_c$, the $\rho/\pi$ abundances are quite different,
because the $\rho$ has a low-mass-estimated to be
$m_\rho^\star=2 m_\pi$ at $T_c+\epsilon$  so that the
$\rho/\pi^-$ ratio measured by STAR is doubled in our scenario \cite{BLR}.
In fact, at $T_c+\epsilon$ our calculated equilibrium abundance
of $\rho$'s is 2-4 times greater than would be found by
equilibration on-shell at $T_c-\epsilon$, where it is usually assumed
in calculating abundances \cite{BMRS}.

We outline above not only what we have done to establish the properties
of the chirally restored mesons, but a whole scenario is laid out for the
experiments now underway at RHIC. Not only will the equilibrium
abundance of the $\rho$-meson be established better by STAR
at $T_c+\epsilon$ (With the present STAR systematic error, the uncertainty is 
a factor of $\sim 2$.), but the dileptons from the decay of the chirally
restored $\rho$ should be seen. Unfortunately, most of them will come in
a wide swath with invariant mass from $0$ to 400 MeV,
just where the ``cocktail" background is largest.
However, the large number should add appreciably to the background, and the
excess from the usual hadronic $\rho$ coming in the mixed phase should be
significantly increased around 400 MeV.

%---------------------------------------------------------------------

\section*{Acknowledgments}
G.E.B. was supported in part by the
US Department of Energy under Grant No. DE-FG02-88ER40388.
C.H.L. was supported by Korea Research Foundation
Grant (KRF-2002-070-C00027) and Pusan National University
Research Grant.

%%%%%%%%%%%%%%%%%%%%%%%%%%%%%%%%%%%%%%%%%%%%%%%%%%%%%%%%%%%%%%%%%%%%%%%%%%%%%%%%

\end{document}